\documentclass[11pt, a4paper]{article}
\usepackage{yfonts}
\usepackage{graphicx}%
\usepackage{multirow}%
\usepackage{amsmath,amssymb,amsfonts}%
\usepackage{amsthm}%
\usepackage{mathrsfs}%
\usepackage[title]{appendix}%
\usepackage{xcolor}%
\usepackage{textcomp}%
\usepackage{manyfoot}%
\usepackage{booktabs}%
\usepackage{algorithm}%
\usepackage{algorithmicx}%
\usepackage{algpseudocode}%
\usepackage{authblk}
\usepackage{listings}%
\usepackage{makecell}
\usepackage{float}
\usepackage{subfig}
\usepackage{wrapfig}
\usepackage{braket}
\usepackage{bbold}
\usepackage{physics}
\usepackage{cancel}
\usepackage{diagbox}
\usepackage{bm}
\usepackage{comment}

\author[1]{Rocco Malaspina}
\author[2,3]{Lorenzo Pierini}
\author[4,5]{Olga Shekhovtsova}
\author[1,5]{Simone Pacetti}

\affil[1]{Dipartimento di Fisica e Geologia, Università degli studi di Perugia, Via Alessandro Pascoli snc, Perugia, 06123, Ital}
\affil[2]{Dipartimento di Fisica e Scienze della Terra, Università degli studi di Ferrara, Via Giuseppe Saragat, 1, Ferrara, 44122, Italy}
\affil[3]{INFN Sezione di Ferrara, 44122, Italy}
\affil[4]{NSC Kharkov Institute for Physics and Technology, Institute for Theoretical Physics, Kharkiv, 61108, Ukraine}
\affil[5]{INFN Sezione di Perugia, 06123, Italy}

\title{Analytical Inverse QCD Coupling Constant approach and its result for $\alpha_s$}

\date{}

\begin{document}
\maketitle
\abstract{We propose a model for the QCD running coupling constant based on the Analytical Inverse QCD Coupling Constant concept with an additional regularization in the low momentum region. Analyticity in the $q^2$-complex plane, where $q$ is the 4-momentum transfer, is imposed by methods of the Analytic Perturbation Theory. The model incorporates a peculiar low-momentum behavior for $\alpha_s(q^2)$ as a divergence at $q^2=0$ to retrieve color confinement, without spoiling its correct high-momentum behavior. This was achieved by means of a two-parameter regularization function, for which we considered three possible analytic expressions. In fact, in the framework of the Analytic Perturbation Theory, $\alpha_s(q^2)$ assumes a finite value for $q^2=0$, at all perturbative orders (\emph{infrared stability}), hence the infrared divergence can not be implemented.
For this reason, we found it more straightforward to work with its reciprocal, namely $\varepsilon_s(q^2) = 1/\alpha_s(q^2)$, imposing its vanishing at the origin of the $q^2$-complex plane via the multiplication of the aforementioned regularizing functions to the spectral density. Once the two free parameters of the regularization functions are settled by fitting to the experimental values of $\alpha_s(q^2)$ at the momenta where these data are available and reliable, the model can reproduce the QCD running coupling constant at any other momentum transferred.}   

\section{Introduction}

The goal of our model is to define and use the inverse QCD coupling constant (ICC)
\begin{equation}
    \varepsilon_s(q^2)=\frac{1}{\alpha_s(q^2)}\,.
    \label{epsilon}
\end{equation}
As we will see in more detail in the following, the advantages of using the ICC go beyond the simpler arithmetical inversion of $\alpha_s(q^2)$ in the formulae~\cite{Srivastava:2001ts}. The failure of perturbation theories at renormalization scales where the running coupling constant is approaching from below the edge of the convergence domain is a well-known limit of such theories. In the leading logarithmic approximation, the expression of the QCD running coupling constant $\alpha_s(q^2)$ is~\cite{Solovtsov_1999,bib1}
\begin{equation}\alpha_s(q^2)=\frac{4\pi}{\beta_0}\frac{1}{\ln \left(-q^2/\Lambda^2\right)}\,, 
\label{eq0} 
\end{equation} 
where $\beta_0=11-2n_f/3$ depends on $n_f$, the number of active quark flavours, i.e., those with masses below the energy $\sqrt{|q^2|}$. The parameter $\Lambda$ represents \emph{Landau's pole} or \emph{ghost pole} at the space-like momentum squared $q^2=-\Lambda^2$. Approximations above the leading order of perturbation theory present similar issues, namely logarithmic divergences.
\\
A procedure adopted to avoid the insurgence of these singularities in the definition of running coupling constant is the Analytic Perturbation Theory (APT)~\cite{bib1,Solovtsov_1999, PhysRevD.55.5295}, which indeed aims to improve the results of Perturbation Theory (PT) in Quantum Field Theories (QFTs), imposing the general principles of analyticity, hence causality, and unitarity. It assumes that propagators and coupling constants, as functions of $q^2$, can be extended analytically in the whole $q^2$-complex plane using the \emph{K\"all\'en-Lehmann’s Spectral Representation} (KL), which formally is a dispersion relation. For the QCD running coupling, the KL is~\cite{Solovtsov_1999}
\begin{equation}
[\alpha_s(q^2)]_{\rm an}= \frac{1}{\pi} \int \limits_0^\infty d\sigma \frac{\rho(\sigma)}{\sigma-q^2}\,,
\label{eq1} 
\end{equation}
where $\rho(\sigma)$ is the \emph{spectral density} and corresponds to the imaginary part of $\alpha_s(q^2)$ calculated on the lower edge of the physical cut, i.e.,
\begin{equation}
\rho(\sigma)=\lim_{\epsilon\to0^+}\mathrm{Im}[\alpha_s(-\sigma-i\epsilon)]\,.
\label{eq2}
\end{equation} 
In this way, all the unphysical singularities produced as artifacts of the PT expansion at finite order, such as ghost poles and unphysical cuts, are eliminated. We worked with the cut along the negative real semi-axis, taking $Q^2=-q^2$ as the opposite of the space-like 4-momentum squared. From Eq.~\eqref{eq0}, the expression of $\alpha_s$ at leading order becomes
\begin{equation}
\alpha_s(Q^2)=\frac{4\pi}{\beta_0}\frac{1}{\ln ({Q^2}/{\Lambda^2})}\,, 
\label{eq3} 
\end{equation} 
where the argument has been simply changed in $Q^2$ omitting the negative sign. Of course, in this form, the Landau's ghost pole occurs at $Q^2=\Lambda^2$. 
\\
By calculating the spectral density $\rho(\sigma)$ from Eq.~\eqref{eq2} using Eq.~\eqref{eq3} and then solving the integral of Eq.~\eqref{eq1}, we obtain 
\begin{equation}
[\alpha_s(Q^2)]_{\rm an}=\frac{4\pi}{\beta_0}\Biggl[\frac{1}{\ln \bigl({Q^2}/{\Lambda^2}\bigr)}+\frac{\Lambda^2}{\Lambda^2-Q^2}\Biggr]\,,\nonumber
\end{equation} 
which is regular at $Q^2=\Lambda^2$. The Landau's pole has been subtracted and hence $[\alpha_s(Q^2)]_{\rm an}$ is finite in the infrared (IR) region, namely at $Q^2\ll\Lambda^2$, in particular, its value in the origin $Q^2=0$ is 
\begin{equation}
[\alpha_s(0)]_{\rm an}=\frac{4\pi}{\beta_0} \,.
\nonumber
\end{equation} 
Moreover, this  result is  independent of the order of the loop expansion~\cite{PhysRevD.55.5295}, a property of the theory called \emph{infrared stability}~\cite{Solovtsov_1999, bib1}. As a consequence, to have a coupling constant with an explosive IR behavior, divergent in the limit $Q^2\to0$, i.e., a behavior which could produce the color confinement phenomenon of QCD, it needs to define a different spectral density. 
\\
The goal of our model is then to go beyond the APT result, which relies on the simple cancellation of Landau’s ghost pole either by subtraction or multiplication. Indeed, in our case, Landau’s pole problem is bypassed by defining a spectral function for the ICC defined in Eq.~\eqref{epsilon}, which is inferred by the leading order expression of the QCD running coupling constant of Eq.~\eqref{eq3}, and by analytically continuating it at each $Q^2$.  The advantage of working with the ICC is consequently explained by the fact that confinement is translated into the going to zero at $Q^2=0$. Also, this condition implies the vanishing of the imaginary part of the ICC, which corresponds to the spectral density of Eq.~\eqref{eq2}, in the same limit $Q^2\to0$.
\\
 Since the APT does not give directly a confining expression of the ICC, we introduce three different types of regularizing function in the spectral density of the KL to incorporate color confinement. 
\\
The paper is organized as follows. In Section 2 we describe the main features of our model which brings together the methods of APT and the regularizing function approach. We propose three parameterizations for the regularizing function, assuring correct high and low momentum behavior and, as a final result, we present analytical formulae for the momentum dependence of both ICC and $\alpha_s(Q^2)$ within 2-loop orders. Numerical predictions of ICC and the QCD running coupling constant at certain momentum points and their comparison with experimental data are obtained and discussed in Section 3. In particular, we computed the ICC at the $Z$-boson mass, i.e., at $q^2=M_Z^2$. Technical details of the calculations are collected in Appendix~\ref{A}.

\section{Application of APT formalism with additional regularizing functions for the ICC}

To apply the KL to the ICC of Eq.~\eqref{epsilon}, it is necessary to start from its PT-expansion~\cite{tesi:l,tesi:r}, from which we obtain the Renormalization Group Equation for the ICC
\begin{equation} 
\frac{\partial }{\partial \ln(\mu^2)}\frac{1}{4\pi\varepsilon_s}=-\frac{\beta_0}{(4\pi\varepsilon_s)^2}-\frac{\beta_1}{(4\pi\varepsilon_s)^3}+\dots\,,\nonumber
\end{equation}
which implies
\begin{equation}
4\pi \frac{\partial \varepsilon_s}{\partial \ln(\mu^2)} =\beta_0+\frac{\beta_1}{4\pi\varepsilon_s}+\dots = \sum_{n=0}^{\infty}\frac{\beta_n}{(4\pi\varepsilon_s)^n}\,.
\label{PT:ICC}
\end{equation}
The solution of the Eq.~\eqref{PT:ICC} truncated at the leading ultraviolet (UV) behavior of the 2-loop order is
\begin{equation}
    \varepsilon_s(Q^2)=\underbrace{K_0\ln\biggl(\frac{Q^2}{\Lambda^2}\biggr)}_{\varepsilon_s^{(0)}}+\underbrace{K_1\ln\biggl(\frac{K_0}{K_1}\ln\Bigl(\frac{Q^2}{\Lambda^2}\Bigr)\biggr)}_{\varepsilon_s^{(1)}}\,,
    \label{eq4}
\end{equation}
where $K_0=\beta_0/(4\pi)$, $K_1=\beta_1/(4\pi\beta_0)$ with $\beta_1=102-38n_f/3$. For $n_f=5$ it is $K_0 \simeq 0.61$ and $K_1 \simeq 0.40$. The 1-loop term is $\varepsilon_s^{(0)}(Q^2)$ while $\varepsilon_s^{(1)}(Q^2)$ is the 2-loop correction. Using the KL, we obtain the analytical expressions
\begin{equation}
\begin{split}
&[\varepsilon_s^{(0,1)}(t)]_{\rm an} =[\varepsilon_s^{(0,1)}(\Lambda^2)]_{\rm an}+\frac{(\Lambda^2-t)}{\pi}\int \limits_0^{\infty}d\sigma \frac{\rho^{(0,1)}(\sigma)}{(\sigma+t)(\sigma+\Lambda^2)}\,,
\end{split}
\label{eq5}
\end{equation} 
where $t=Q^2$ is the space-like Mandelstam variable. The dispersion relation subtracted at $t=\Lambda^2$ is required, otherwise the integral would be divergent for any complex value of $t$. The expressions of the two spectral densities are obtained from
\begin{equation} 
\rho^{(0,1)}(\sigma)=\lim_{\alpha\to0^+}\text{Im}[\varepsilon_s^{(0,1)}(-\sigma-i\alpha)]\,.
\label{rho:eps}
\end{equation}
Inserting the two terms of Eq.~\eqref{eq4} in Eq.~\eqref{rho:eps} we have
\begin{align}
    &\rho^{(0)}(\sigma)=-\pi K_0\,,
    \nonumber
    \\
    &\rho^{(1)}(\sigma)= -K_1 \mathrm{arccotan}\left(\frac{\ln(\sigma/\Lambda^2)}{\pi}\right)\,. \label{eq7}
\end{align}
Evaluating the two integrals of Eq.~\eqref{eq5} we have the following expressions
\begin{align}
    &[\varepsilon_s^{(0)}(t)]_{\rm an}=K_0\ln\left(\frac{t}{\Lambda^2}\right)\,,
    \label{eq8}\\ 
    &[\varepsilon_s^{(1)}(t)]_{\rm an}=K_1\ln\left(\frac{t}{t-\Lambda^2}\ln\left(\frac{t}{\Lambda^2}\right)\right)\,.\label{eq9}
\end{align}
Similarly to $[\alpha_s(q^2)]_{\rm an}$, Eqs.~\eqref{eq8} and~\eqref{eq9} have the correct analytical behavior, although they do not show confinement. We then introduce a \emph{regularizing function} $r(\sigma)$ in the spectral density to achieve
\begin{equation}
\lim_{Q^2\to0^+}[\varepsilon_s(Q^2)]_{\rm an}=0\,.
\nonumber
\end{equation}
This means that also the spectral density, which is the imaginary part of the ICC, has to be zero in the same limit. Moreover, the regularizing function must also not spoil the correct perturbative UV limit. Therefore, this function can be chosen from the set of arbitrary continuous functions fulfilling the conditions
\begin{equation}
r(\sigma)
\begin{cases}
 \displaystyle\mathop{\longrightarrow}_{\sigma\to0^+} 0\\
 \displaystyle\mathop{\longrightarrow}_{\sigma\to +\infty}1
\end{cases}\,.
\label{eq10}
\end{equation}
It follows that the KL representation becomes
\begin{equation}
\begin{split}
&[\overline{\varepsilon}_s^{(0,1)}(t)]_{\rm an} =[\overline{\varepsilon}_s^{(0,1)}(\Lambda^2)]_{\rm an}+\frac{(\Lambda^2-t)}{\pi}\int \limits_0^{\infty}d\sigma \frac{\overline{\rho}^{(0,1)}(\sigma)}{(\sigma+t)(\sigma+\Lambda^2)}\,,%
\end{split}
\label{eq11}
\end{equation}
where the regularized spectral densities are
\begin{equation}
\overline{\rho}^{(0,1)}(\sigma)=\rho^{(0,1)}(\sigma) r(\sigma)\,. \nonumber
\end{equation}
\begin{figure}[]%
\begin{center}
\includegraphics[width=0.8\textwidth]{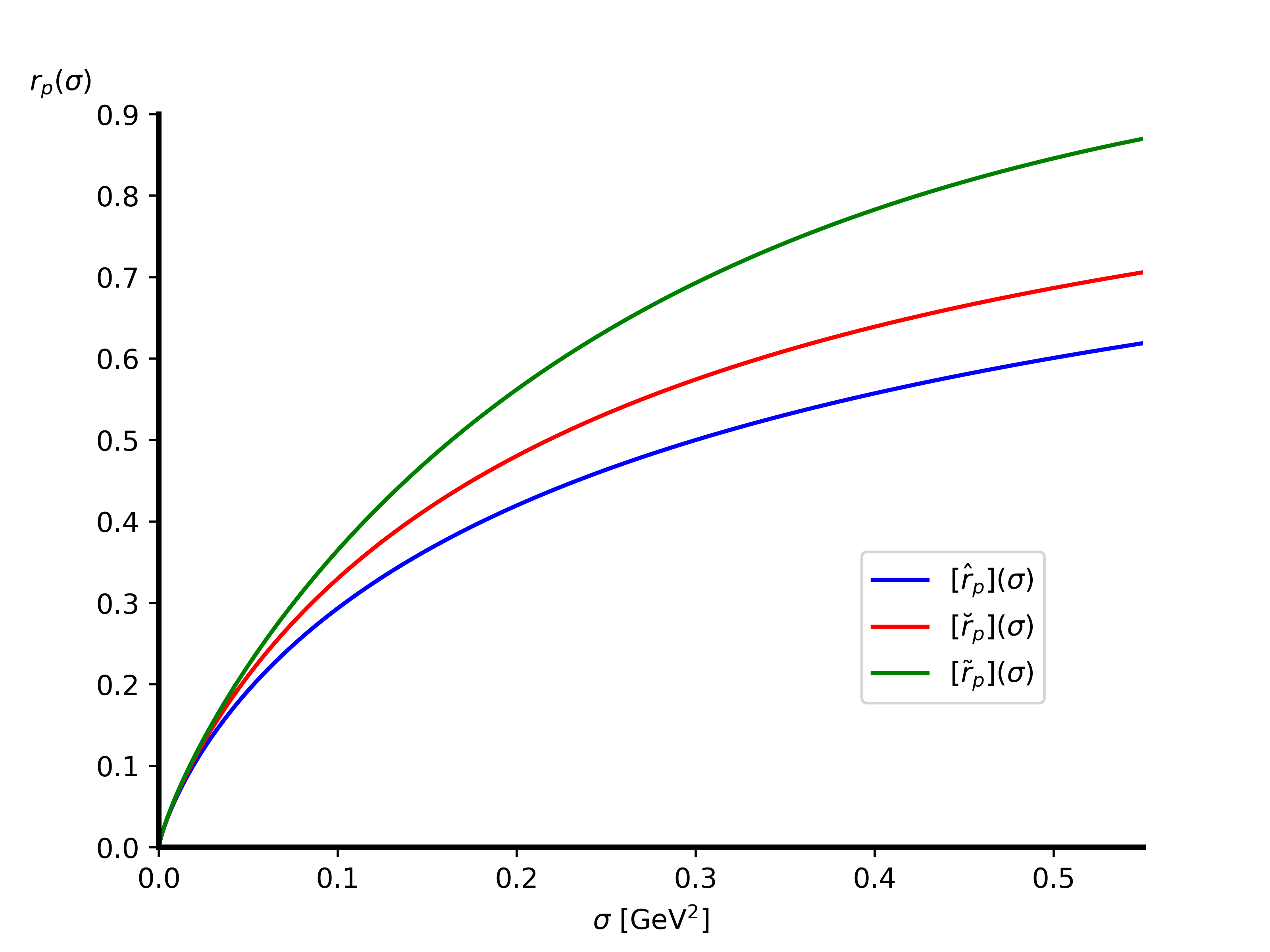}
\caption{Momentum spectra for the three regularizing functions, $p=0.8$ and $\Lambda=300 \,\, \textrm{MeV}$.}
\label{fig1}
\end{center}
\end{figure}
The subtraction parameter $[\overline{\varepsilon}_s^{(0,1)}(\Lambda^2)]_{\rm an}$ is then fixed by requiring the vanishing at $Q^2=0$ of the functions of Eq.~\eqref{eq11}. The three parameterizations for the regularizing function  $r(\sigma)$ assuring the correct high and low-momentum behavior of Eq.~\eqref{eq10} are
\begin{align}
&\Hat{r}_p(\sigma)=\frac{1}{1+\left({\Lambda^2}/{\sigma}\right)^p}\,;\label{eq12_Pacetti}\\
&\Breve{r}_p(\sigma)=\frac{1}{\left(1+{\Lambda^2}/{\sigma}\right)^p}\,;\label{eq13_Rocco}\\
&\Tilde{r}_p(\sigma)=\left(1-e^{-{\sigma}/{\Lambda^2}}\right)^p \,, \label{eq14_Lorenzo}
\end{align}
where $p \in (0,1)$ and $\sigma\in\mathbb{R}^+$. The constraints on $p$ are demanded by Eq.~\eqref{eq10} and, in addition, the condition $p<1$ guarantees the convergence of some integrals appearing in the \emph{soft gluon resummation} theory~\cite{bib2}. In other words, we want $[\overline \alpha_s(q^2)]_{\rm an}$ to be divergent in the IR region but still integrable.
\\
To simplify the reading, we present here the scheme to get $[\overline{\varepsilon}_s^{(0,1)}(t)]_{\rm an}$ only for the regularizing function in the form of Eq. \eqref{eq14_Lorenzo}. In a similar way $[\overline{\varepsilon}_s^{(0,1)}(t)]_{\rm an}$ can be obtained for the other two parameterizations. The corresponding calculations are reported in the Appendix~\ref{A}.
\\
\\
So, following Eq.~\eqref{eq4}, the zero-order ICC is written as
\begin{equation}
\begin{split}
[\Tilde{\overline{\varepsilon}}_s^{(0)}(t)]_{\rm an}=&[\Tilde{\overline{\varepsilon}}_s^{(0)}(\Lambda^2)]_{\rm an}+\frac{\Lambda^2-t}{\pi}\int_0^{\infty}d\sigma \frac{\Tilde{\overline{\rho}}^{(0)}(\sigma)}{(\sigma+t)(\sigma+\Lambda^2)}\,,
\end{split}\nonumber
\end{equation}
where the regularized spectral density is
\begin{equation}
\Tilde{\overline{\rho}}^{(0)}(\sigma)=\rho^{(0)}(\sigma) \Tilde{r}_p(\sigma)=-\pi K_0 \left(1-e^{-{\sigma}/{\Lambda^2}}\right)^p\,.\nonumber
\end{equation}
Using the expression of Eq.~\eqref{eq14_Lorenzo} for the spectral density we find the following form
\begin{equation}
[\Tilde{\overline{\varepsilon}}_s^{(0)}(z)]_{\rm an}=[\Tilde{\overline{\varepsilon}}_s^{(0)}(\Lambda^2)]_{\rm an}-K_0 F(z)\,,
	\nonumber
\end{equation}
where
\begin{equation}
F(z)=(1-z)\int_0^{\infty}dx \frac{\bigl(1-e^{-x}\bigr)^p}{(x+z)(x+1)}\,.
\nonumber
\end{equation}
By means of the expansion for the regularizing function
\begin{equation}
\Tilde{r}_p(x)=\bigl(1-e^{-x}\bigr)^p
=\sum_{k=0}^{\infty}(-1)^k\binom{p}{k}e^{-kx}\,,
\nonumber
\end{equation}
with the convergence condition $e^{-x}\leq 1$, we obtain
\begin{equation}
\begin{split}
&[\Tilde{\overline{\varepsilon}}_s^{(0)}(z)]_{\rm an}=[\varepsilon_s^{(0)}(z)]_{\rm an}+\gamma K_0\sum_{k=1}^{\infty}(-1)^k\binom{p}{k}\left[e^{kz}\Gamma(0,kz)+\ln(k)\right]\,,
\end{split}
\label{eq20}
\end{equation}
where $\gamma\simeq 0.57721$ is the Euler-Mascheroni's constant and
\begin{equation}
\Gamma(0,x)=\int_{x}^{\infty}\frac{e^{-t}}{t}dt=E_1(x)\,.
\nonumber
\end{equation}
is the {\it exponential integral function}~\cite{tesi:l}.
\\
\\
We have calculated an expression for the ICC at the zeroth order, which is given by a sum of the analytical unconfined ICC of Eq.~\eqref{eq8} and several confining corrections terms, whose definitions contain the function $\Gamma(0,x)=E_1(x)$ and the constant $\gamma$.
\\
At the same time, the confined expression for ICC at the first order is obtained repeating the same steps of Eq.~\eqref{eq20}. In this way we obtain an expression which is written in the form of confining corrections to the analytical unconfined contribution of ICC at the first order, see Eq.~\eqref{eq9}, but in this case the confining corrections are written in an implicit integral form, i.e.,
\begin{equation}
    [\Tilde{\overline{\varepsilon}}_s^{(1)}(z)]_{\rm an}=[\varepsilon_s^{(1)}(t)]_{\rm an}
    +\sum_{k=1}^{\infty}(-1)^k\binom{p}{k}\Tilde{I}_k(z)\,,
\label{eq21}
\end{equation}
with
\begin{equation}
\begin{split}
    \Tilde{I}_k(z)=&\frac{1}{\pi}\int_0^{\infty}dx \frac{\rho^{(1)}(x)(1-e^{-kx})}{x(x+1)}+\frac{1-z}{\pi}\int_0^{\infty}dx \frac{\rho^{(1)}(x)\cdot e^{-kx}}{(x+z)(x+1)}\,.
\end{split}
\label{integral_1}
\end{equation}
These integrals are not solvable in a closed form because of the function $\rho^{(1)}(x)$, given in Eq.~\eqref{eq7}. Some procedures for solving the integrals of Eq.~\eqref{integral_1} are defined in Ref.~\cite{tesi:l}.
\\
The details regarding the calculations of the other two parameterizations are given in the Appendix~\ref{A} and additional details can be found in Refs.~\cite{tesi:l, tesi:r}.
\\
Our main analytical result can be summarized by the expression
\begin{equation}
[\overline{\varepsilon}_s(t)]_{\rm an}=[\overline{\varepsilon}_s^{(0)}(t)]_{\rm an}+[\overline{\varepsilon}_s^{(1)}(t)]_{\rm an} \nonumber\,, 
\end{equation}
where $\overline{\varepsilon}_s^{(0,1)}$ for the regularizing function with the exponential of Eq.~\eqref{eq14_Lorenzo} are presented in Eqs.~\eqref{eq20} and~\eqref{eq21} and the remaining two parameterizations are presented in the Appendix~\ref{A}.

\section{Numerical results}

\begin{figure}[]%
\begin{center}
\includegraphics[width=0.8\textwidth]{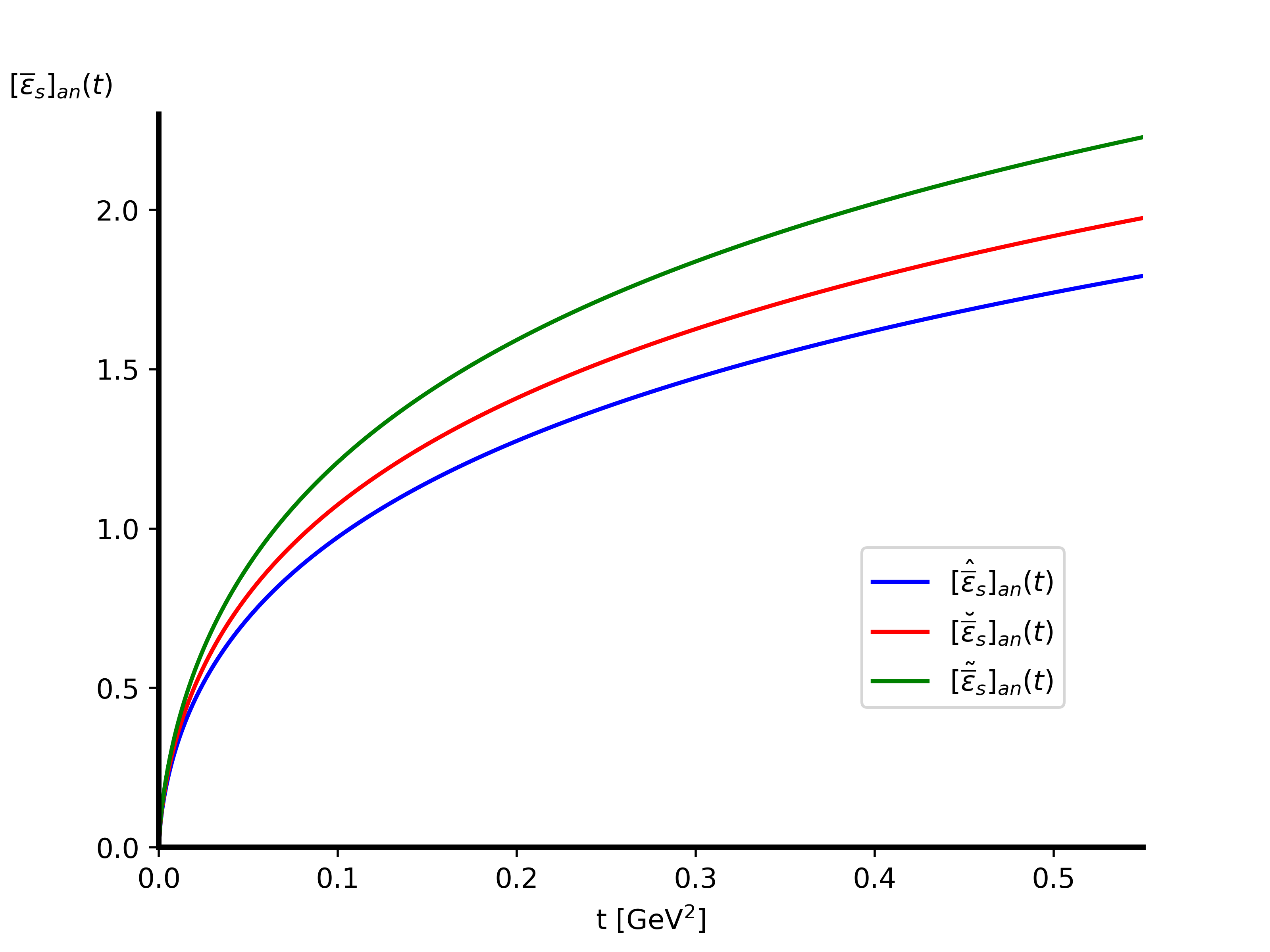}
\caption{4-momentum squared distribution for  $[\overline{\varepsilon}_s(t)]_{\rm an}$,  $p=0.8$, $\Lambda=300 \,\, \textrm{MeV}$ and $n_f=5$.}
\label{fig2}
\end{center}
\end{figure}

\begin{figure}[]%
\begin{center}
\includegraphics[width=0.8\textwidth]{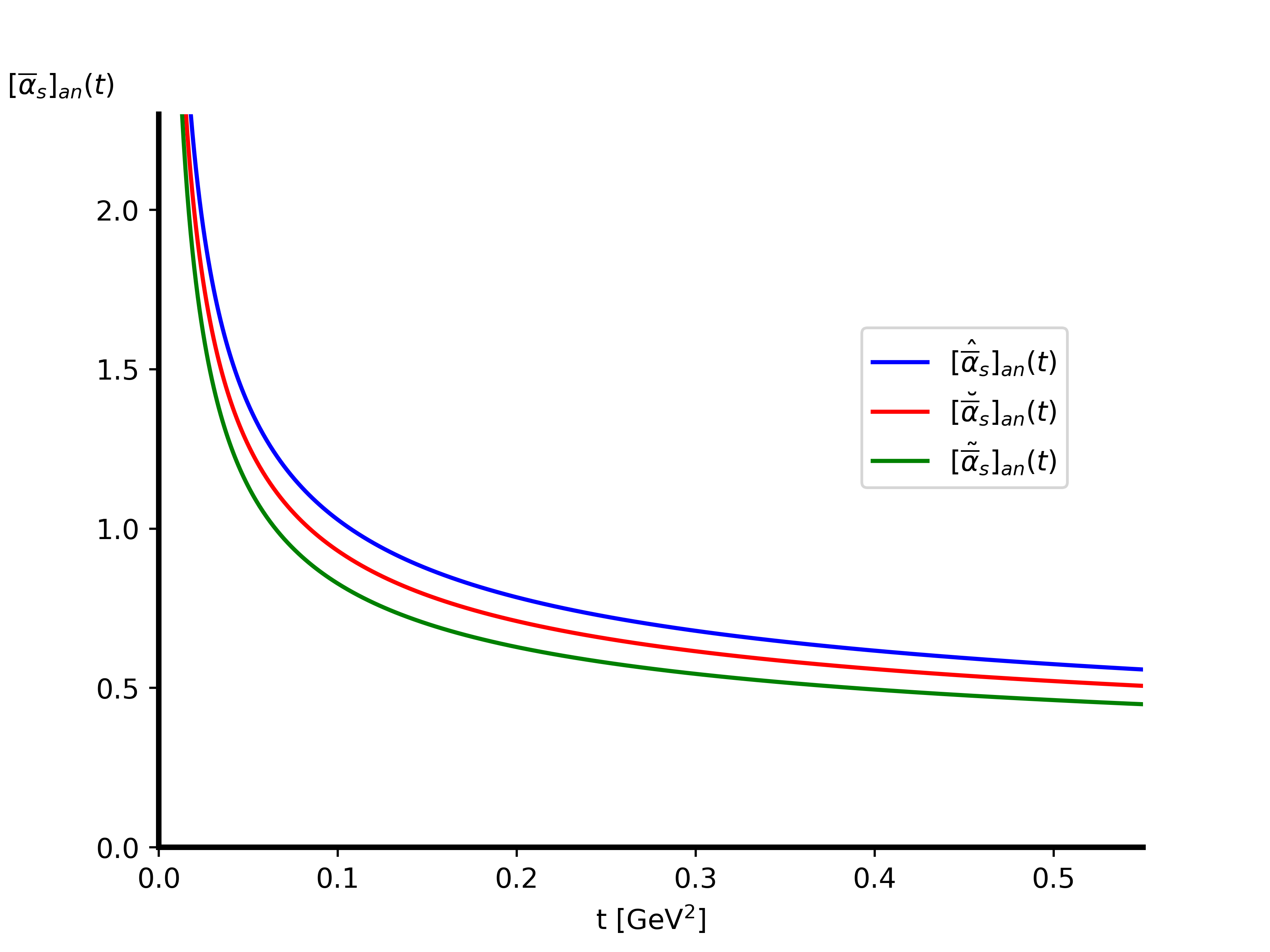}
\caption{4-momentum squared distribution for $[\overline{\alpha}_s(t)]_{\rm an}$, for $p=0.8$, $\Lambda=300 \,\, \textrm{MeV}$ and $n_f=5$.}
\label{fig3}
\end{center}
\end{figure}

First, we will illustrate the results obtained in the previous sections by considering the momentum spectra for both $[\overline{\varepsilon}_s(t)]_{\rm an}$ and $\alpha_s(t) = 1/[\overline{\varepsilon}_s(t)]_{\rm an}$.
\\
In Figs.~\ref{fig2} and \ref{fig3} we show the momentum distributions according to our model for three regularizing function parametrizations, where $n_f=5$ and  $p=0.8$, $\Lambda=300 \,\, \textrm{MeV}$. 

As a next step  we calculate $\alpha_s$ at the $Z$-boson mass. Results for several values of $p$ and $\Lambda$ are reported in the Tables~\ref{tab1},~\ref{tab2} and~\ref{tab3}. In fact, our values $\alpha_s(M_Z^2)$ are compatible with the world average value~\cite{huston2023quantum}\\
\begin{equation}
\alpha^{\text{PDG}}
_s(M_Z^2)=0.1180\pm0.0009 \,,
\nonumber
\end{equation}
for all three regularizing function parameterizations. For all values of $p$ and $\Lambda$ presented in  the Tables~\ref{tab1},~\ref{tab2} and~\ref{tab3} the difference between the model prediction and the PDG value is less than $10\%$ for all three parameterizations.

\newpage

\begin{table}[h]
\begin{center}
\begin{tabular}{| c | c | c | c | c |}
    \hline
       & \multicolumn{4}{c|}{\Gape[0.15cm][0.15cm]{$[\Hat{\overline{\alpha}}_s]_{\rm an}$}} \\ 
   \hline
 \diagbox{$\Lambda$}{$p$} & 0.5 & 0.6 & 0.7 & 0.8\\
    \hline
0.2   &   0.11612  &   0.11683  &   0.11729  &   0.11761  \\
    \hline
0.3   &  0.12358    &  0.12441     &  0.12494     &   0.12530  \\
    \hline
0.4   &   0.12949     &    0.13043   &   0.13102 & 0.13142  \\
    \hline
\end{tabular}
\captionof{table}{Numerical results for $[\Hat{\overline{\alpha}}_s(M_Z^2)]_{\rm an}$.}
\label{tab1}
\end{center}
\end{table}\vspace{-10mm}

\begin{table}[h]
\begin{center}
\begin{tabular} {| c | c | c | c | c |}
    \hline
            & \multicolumn{4}{c|}{\Gape[0.15cm][0.15cm]{$[\Breve{\overline{\alpha}}_s]_{\rm an}$}} \\
   \hline
 \diagbox{$\Lambda$}{$p$} & 0.5 & 0.6 & 0.7 & 0.8 \\
    \hline
0.2  &   0.10342  &   0.10756 &   0.11087    &  0.11362 \\
    \hline
0.3   &   0.10932    &  0.11396   &   0.11769    &   0.12079  \\
    \hline
0.4  &  0.11395     & 0.11900    &   0.12307    &    0.12647  \\
    \hline
\end{tabular}
\captionof{table}{Numerical results for the model $[\Breve{\overline{\alpha}}_s(M_Z^2)]_{\rm an}$.}
\label{tab2}
\end{center}
\end{table}\vspace{-10mm}
\begin{table}[h]
\begin{center}
\begin{tabular} {| c | c | c | c | c |}
    \hline & \multicolumn{4}{c|}{\Gape[0.15cm][0.15cm]{$[\Tilde{\overline{\alpha}}_s]_{\rm an}$}} \\
   \hline
 \diagbox{$\Lambda$}{$p$} & 0.5 & 0.6 & 0.7 & 0.8 \\
    \hline
0.2  &  0.10058     &   0.10406    &   0.10674    &  0.10887 \\
    \hline
0.3   &   0.10616    &  0.11004     &   0.11304    &  0.11544 \\
    \hline
0.4 & 0.11052   &   0.11473    & 0.11800     &  0.12062  \\
    \hline
\end{tabular}
\captionof{table}{Numerical results for the model $[\Tilde{\overline{\alpha}}_s(M_Z^2)]_{\rm an}$.}
\label{tab3}
\end{center}
\end{table}

\begin{figure}[]
\begin{center}
	\includegraphics[width=0.7\textwidth]{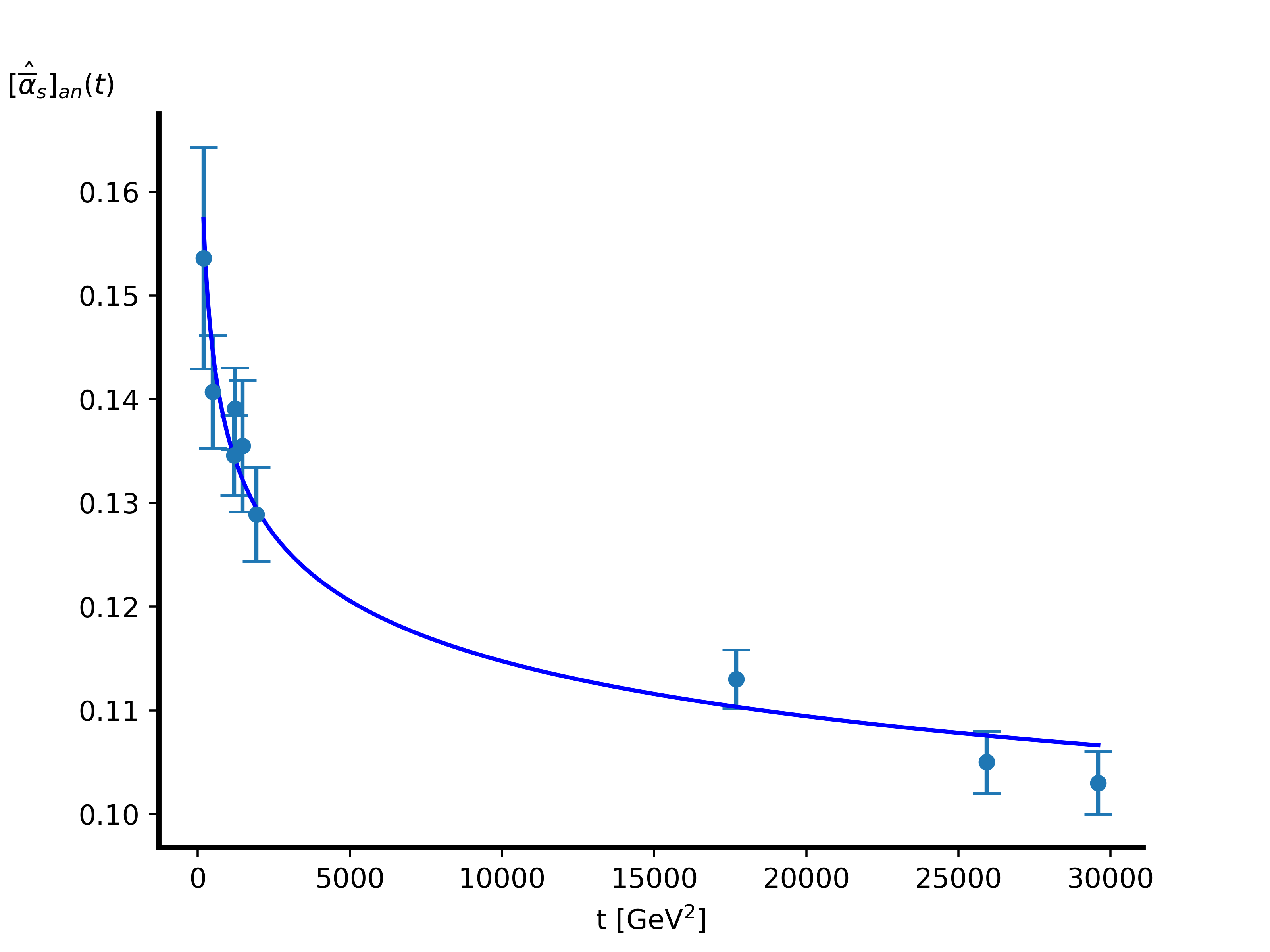}
	\includegraphics[width=0.7\textwidth]{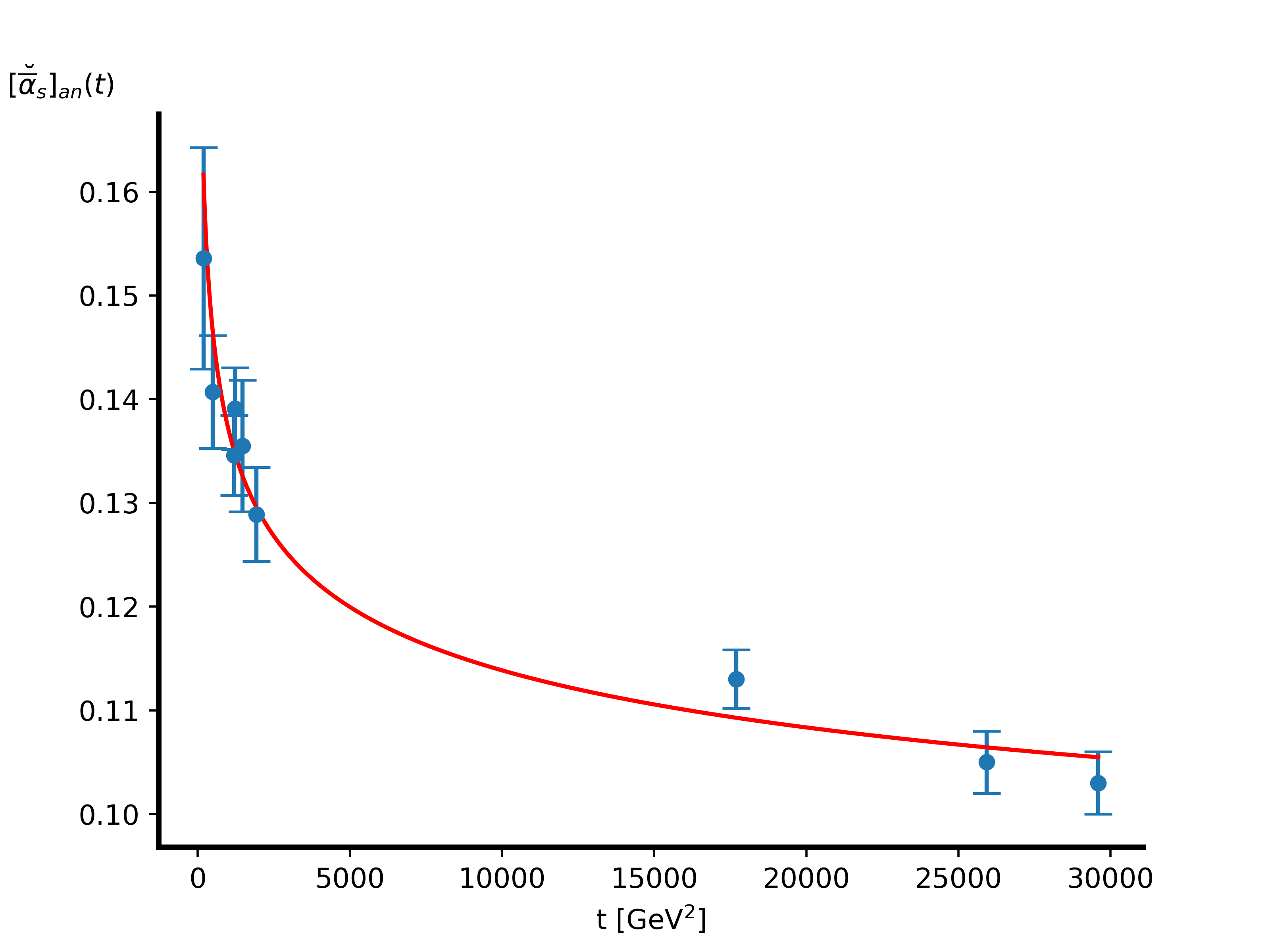}	
	\includegraphics[width=0.7\textwidth]{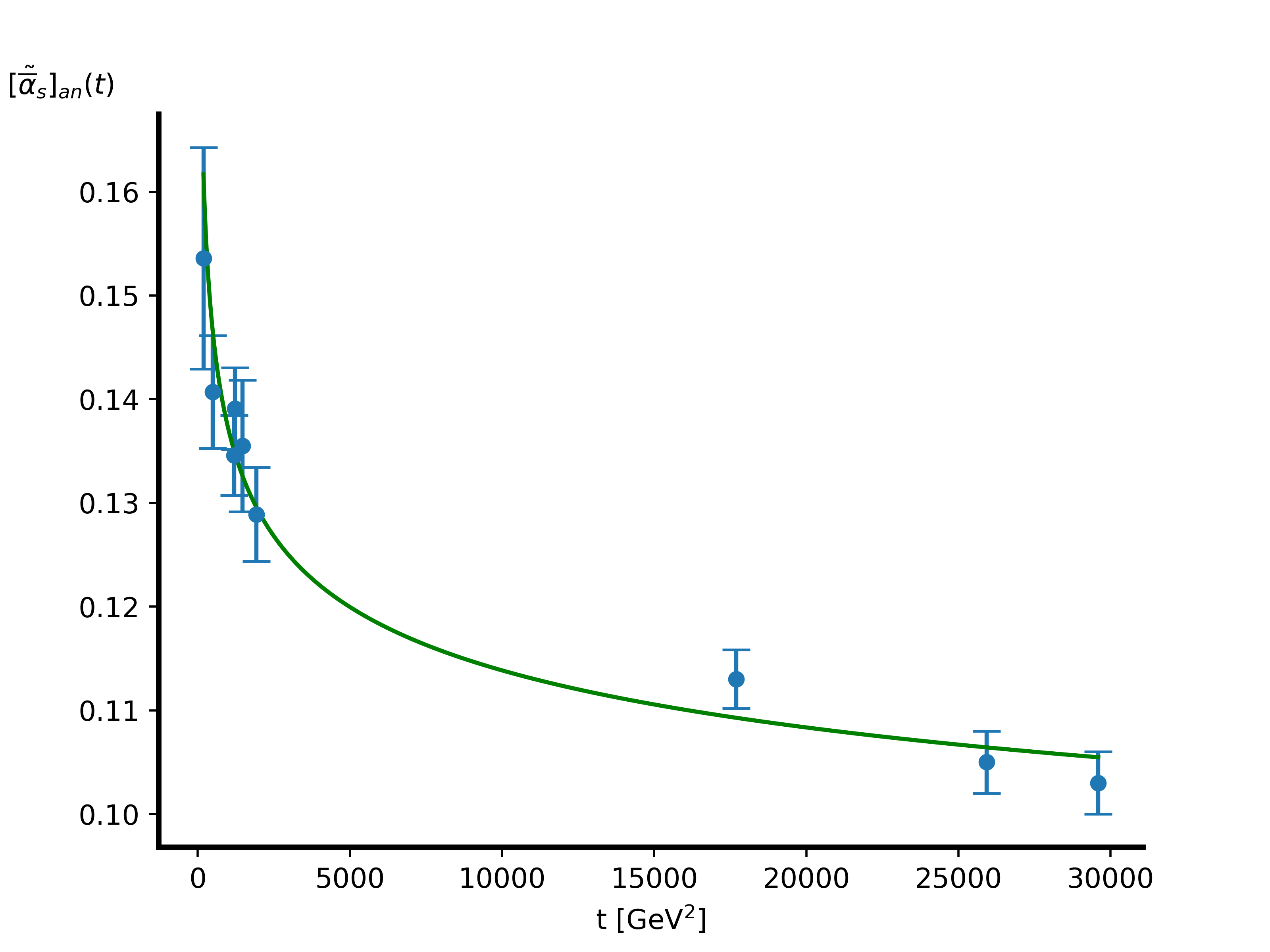}
	\caption{Optimal curves for momentum evolution of the running coupling constant ${\overline{\alpha}}_s(t)$,  $\Lambda=300$ MeV.}
	\label{fit:fig}
\end{center}
\end{figure}

By setting $\Lambda = 300$ MeV, the values of the parameter $p$ obtained fitting to the data for $\alpha_s$, measured by the experiments JADE, LEPII and CMS~\cite{Khachatryan_2015} and the corresponding $\chi^2$'s are
\begin{equation}
\begin{split}
&[\Hat{\overline{\alpha}}_s]_{\rm an}: \quad p=0.25 \pm 0.01 \quad (\chi^2=0.695)\,;\\
&[\Breve{\overline{\alpha}}_s]_{\rm an} : \quad p=0.64 \pm 0.03 \quad (\chi^2=0.723)\,;\\
&[\Tilde{\overline{\alpha}}_s]_{\rm an}: \quad p= 0.80 \pm 0.05 \quad (\chi^2=0.724)\,. 
\end{split}\nonumber
\end{equation}
The theoretical curves are shown in Fig.~\ref{fit:fig}.
\\
In conclusion, it is worth mentioning that all results presented here can be produced using our codes~\cite{git:link}, both Mathematica~\cite{Mathematica} and Python versions are presented there. The numerical values produced by the codes differ for less then $0.01\%$ and our Tables~\ref{tab1},~\ref{tab2} and~\ref{tab3} contain the values obtained from the Python code. It is to be noted that the 'quad' method from the scientific library SciPy~\cite{scipy} was applied in Phython code whereas the "GaussKronrodRule" method was chosen in Mathematica.

\section{Conclusions}
This study represents an attempt to define an analytic model for the QCD coupling constant, where, for the first time, the function of interest is the ICC. The advantage of using the inverse of the coupling constant, which could be interpreted as the QCD vacuum permittivity, lies in the possibility of formally treating poles and hence singularities of the coupling constant, since they coincide with zeros of the ICC. It follows that, assuming quite naturally that the running coupling constant does not vanish at finite, the analyticity domain of the ICC is larger than that of $\alpha_s(Q^2)$. In particular, the phenomenon of the QCD color confinement occurring at low momentum, i.e., at large distances, assumed as a consequence of the divergence of the coupling constant as $Q^2$ goes to zero, should correspond to a regular zero for the ICC in the same limit $Q^2\to 0$.
\\
The procedure for defining an analytic expression of the ICC as a function of the 4-momentum transferred squared, having the desired IR behavior, has been achieved by exploiting the APT approach to introduce a parametric regularizing function which assured the vanishing of the ICC as $Q^2\to0$.
\\
Three possible parameterizations have been considered for such a regularizing function. All the three expressions, given in Eqs.~\eqref{eq12_Pacetti},~\eqref{eq13_Rocco} and~\eqref{eq14_Lorenzo}, depend on the same pair of parameters, namely the adimensional power $p$, whose values are limited in the interval $(0,1)$ by the convergence condition of the dispersion-relations integral, and the momentum scale $\Lambda$. 
\\
Concerning the meaning of these parameters, while the $\Lambda$ can be naturally identified as the QCD momentum scale, the adimensional $p$ instead does not have a clear physical interpretation. There are, however, studies, see e.g.  Ref.~\cite{bib2}, where a similar $p$-power law has been used to regularize soft-gluon resummation in the IR limit of QCD.
\\
Finally, we have presented the first attempts to fit the free parameters of the regularizing functions to the data on the QCD coupling constant. Preliminary results, shown in the three panels of Fig.~\ref{fit:fig}, which correspond to the three parameterizations of the regularizing functions given in Eqs.~\eqref{eq12_Pacetti},~\eqref{eq13_Rocco} and~\eqref{eq14_Lorenzo}, are quite encouraging because the values which provide the best description of the data are in agreement with the physical expectations. In these cases, by setting the momentum scale $\Lambda$ to the value of 300 MeV, only single-parameter fits have been performed. Such a limitation has been due to the too-high computing power required by the double-parameter fit procedure.
\\
A complete study is in progress where the effect of both parameters is taken into account, also considering other observables in which the $Q^2$-functional form of $\alpha_s$, especially in the IR region, plays a crucial role, such as the hadronic contributions~\cite{Eidelman_1995, WorkingGrouponRadiativeCorrections:2010bjp}  to the anomalous magnetic moment of the muon~\cite{PhysRevLett.131.161802} and to the inclusive decay width of the $\tau$ lepton~\cite{PICH201441}.


\appendix 
\section{Analytical results for $\varepsilon(t)_{an}$ within APT with regularizing functions}\label{A}
This appendix provides a brief guide for $[\overline{\varepsilon}_s(t)]_{\rm an}$ calculation taking into account the regularizing functions of Eqs.~(\ref{eq12_Pacetti}) and (\ref{eq13_Rocco}).
\subsection{Model $\Hat{r}_p(\sigma)$}
\label{model1}
The parametrization $\Hat{r}_p(\sigma)$ consists in Eq.~\eqref{eq12_Pacetti}. It has to be inserted in the integral of Eq.~\eqref{eq5} to obtain the contributions of both the ICC contributions. The zero order term of the regularized spectral density is
\begin{equation}
\Hat{\overline{\rho}}^{(0)}(\sigma)=\rho^{(0)}(\sigma) \Hat{r}_p(\sigma)=\frac{-\pi K_0}{1+\left({\Lambda^2}/{\sigma}\right)^p}\,.\nonumber
\end{equation}
Inserting this expression in the integral of Eq.~\eqref{eq5} we obtain
\begin{equation}
\begin{split}
    &[\Hat{\overline{\varepsilon}}_s^{(0)}(t)]_{\rm an}=[\Hat{\overline{\varepsilon}}_s^{(0)}(\Lambda^2)]_{\rm an} -K_0\left(\Lambda^2-t\right)\int\limits_0^{\infty}\frac{1}{1+\left(\Lambda^2/\sigma\right)^p}\frac{d\sigma}{\left(\sigma+\Lambda^2\right)\left(\sigma+t\right)}\,.
\end{split}
\nonumber
\end{equation}
In the case of $p=n/m\in\mathbb{Q}$, with $n$ and $m$ positive integers, the result is
\begin{equation}
    \begin{split}
    &[\Hat{\overline{\varepsilon}}_s^{(0)}(t)]_{\rm an}=K_0\ln\left(\frac{t}{\Lambda^2}\right) \left[1+\frac{1}{n}\sum_{j=1}^n\sum_{l=1}^m\frac{\frac{c_l}{m}+\frac{\ln\left(b_j/c_l\right)}{\ln\left(t/\Lambda^2\right)}b_j\left(\frac{t}{\Lambda^2}\right)^{1/m}}{\left(\frac{t}{\Lambda^2}\right)^{1/m}b_j-c_l}\right]\,,
\end{split}\nonumber
\end{equation}
where
\begin{equation}
\begin{cases}
&b_j=\displaystyle-\exp\left(i\pi\frac{2j-1}{n}\right)\\
&\\
&c_l=\displaystyle-\exp\left(i\pi\frac{2l-1}{m}\right)\\
\end{cases}\,,\nonumber
\end{equation}
with $(j,l)\in\{1,2,\ldots,n\}\times\{1,2,\ldots,m\}$.\\ 
Using the spectral density
\begin{equation}
\begin{split}
    \Hat{\overline{\rho}}^{(1)}(\sigma)&=\rho^{(1)}(\sigma) \Hat{r}_p(\sigma)=\frac{-K_1 \mathrm{arccotan}\left(\ln\left(\sigma/\Lambda^2 \right)/{\pi} \right)}{1+\left({\Lambda^2}/{\sigma}\right)^p} \,
\end{split}\nonumber
\end{equation}
we calculated the 2-loop ICC term
\begin{equation}
\begin{split}
    [\Hat{\overline{\varepsilon}}_s^{(1)}(t)]_{\rm an} &=-\frac{K_1}{n}\sum_{j=1}^n\sum_{l=1}^m\frac{b_j}{b_j-z^{-k1/m}c_l} \sum_{k=0}^{\infty}\Biggl[ \ln\left(\frac{\left(y_k^{(j)}-i\pi/m\right)\left(y_k^{(l)}+i\pi/m\right)}{\left(y_k^{(j)}+i\pi/m\right)\left(y_k^{(l)}-i\pi/m\right)}\right)\\
    &+\ln\left(\frac{\left(e^{i\pi/m}+z^{-1/m}c_l\right)\left(1+b_j\right)}{\left(e^{i\pi/m}+b_j\right)\left(1+z^{-1/m}c_l\right)}\right)\Biggr]\,,
\end{split}\nonumber
\end{equation}
where $z=t/\Lambda^2$ and
\begin{equation}
\begin{cases}
    &y_k^{(j)}=\displaystyle\left(\frac{2j-1}{n}+2k\right)i\pi\\
    &\\
    &y_k^{(l)}=\displaystyle\left(\frac{2l-1}{m}+2k\right)i\pi-\frac{\ln(z)}{m}\\
\end{cases}\,\nonumber
\end{equation}
with $(j,l)\in\{1,2,\ldots,n\}\times\{1,2,\ldots,m\}$ and $\forall\,k\in~\mathbb{Z}$.
\subsection{Model $\Breve{r}_p(\sigma)$}
\label{model2}
The parametrization proposed in Eq.~\eqref{eq13_Rocco} gives to the spectral densities the forms~\cite{tesi:r}
\begin{align}
\Breve{\overline{\rho}}^{(0)}(\sigma) &=\frac{-\pi K_0}{\left(1+{\Lambda^2}/{\sigma}\right)^p}, \nonumber
\\ \Breve{\overline{\rho}}^{(1)}(\sigma) &=\frac{-K_1 \mathrm{arccotan}\left({\ln\left(\sigma/\Lambda^2\right)}/{\pi}\right)}{\left(1+{\Lambda^2}/{\sigma}\right)^p}\,.
\nonumber
\end{align}
The two contributions to the ICC are therefore calculated inserting these expressions in Eq.~\eqref{eq5}. 
We get 
\begin{equation}
[\Breve{\overline{\varepsilon}}_s^{(0)}(t)]_{\rm an}=\frac{K_0}{p} {}_2F_1\left(1,p;1+p;\frac{t-\Lambda^2}{t}\right)\,,\nonumber
\end{equation} 
where ${}_2F_1(a,b;c;z)$ is the \emph{Gaussian Hypergeometric Function}, the analytic continuation of the \emph{Gaussian Hypergeometric series}
\begin{equation}
	 \sum_{n=0}^\infty \frac{(a)_n (b_n)}{(c)_n n!} z^n\,,
	 \nonumber
\end{equation}
the Pochhammer symbol here is used for the ascending factorial
\begin{equation}
(a)_n=\frac{\Gamma(a+n)}{\Gamma(a)}\,.\nonumber%
\end{equation}
 The 2-loop ICC term has not been calculated in a closed form, but we arrived at a partial expression which contains a series instead of an integral, valid if $|t|>\Lambda^2$
\begin{equation}
\begin{aligned} 
[{\Breve{\overline{\varepsilon}}_s^{(1)}(t)}]_{\rm an} 
&=K_1 \frac{\sin(\pi p)}{\pi} A + K_1 \left(\frac{t}{t-\Lambda^2} \right)^{p}\ln \left(\ln \frac{t}{\Lambda^2} \right) + K_1 \frac{\pi \cos(\pi p)}{\sin(\pi p)} \left(\frac{t}{t-\Lambda^2}\right)^{p} \\ &+ K_1 \frac{\sin(\pi p)}{\pi} \sum_{l,k=0}^\infty \frac{(p)_k}{k!} \frac{\gamma+\ln(p+k+l+1)}{p+k+l+1} \left(\frac{\Lambda^2}{t}\right)^{l+1} \,,
\end{aligned} 
\nonumber
\end{equation}
where $A$ is a constant given in terms of an integral depending only on the parameter $p$, namely
\begin{equation}
A=-\int \limits_0^1 dy \frac{\ln(- \ln y)}{y} \left(\frac{y}{1-y}\right)^{p}\,. 	\nonumber
 \end{equation}

\bibliographystyle{unsrt}
\bibliography{references.bib}

\end{document}